# Polarized and Unpolarized Nucleon Structure Functions from Lattice QCD

M. Göckeler[1,2], R. Horsley[1,3], E.-M. Ilgenfritz[3], H. Perlt[4], P. Rakow[5], G. Schierholz[6,1] and A. Schiller[4]

[1] Höchstleistungsrechenzentrum HLRZ, c/o Forschungszentrum Jülich,
D-52425 Jülich, Germany

[2] Institut für Theoretische Physik, RWTH Aachen,
D-52056 Aachen, Germany

[3] Institut für Physik, Humboldt-Universität,
D-10115 Berlin, Germany

[4] Fakultät für Physik und Geowissenschaften, Universität Leipzig,
D-04109 Leipzig, Germany

[5] Institut für Theoretische Physik, Freie Universität,
D-14195 Berlin, Germany

[6] Deutsches Elektronen-Synchrotron DESY,
D-22603 Hamburg, Germany

**Abstract**

We report on a high statistics quenched lattice QCD calculation of the deep-inelastic structure functions $F_1$, $F_2$, $g_1$ and $g_2$ of the proton and neutron. The theoretical basis for the calculation is the operator product expansion. We consider the moments of the leading twist operators up to spin four. Using Wilson fermions the calculation is done for three values of $\kappa$, and we perform the extrapolation to the chiral limit. The renormalization constants, which lead us from lattice to continuum operators, are calculated in perturbation theory to one loop order.

# 1 Introduction

Deep-inelastic lepton-nucleon scattering is described by four structure functions: $F_1$, $F_2$, $g_1$ and $g_2$. The spin-averaged structure functions $F_1$ and $F_2$ carry information about the overall density of quarks and gluons in the nucleon. They have played a seminal role in the development of our current understanding of the structure of hadrons. The polarized structure function $g_1$ goes one step further and probes the distribution of quarks of a given helicity in the longitudinally polarized nucleon. Recent measurements of $g_1$ [1] have revealed the (at first sight) surprising result that only a small fraction of the nucleon's spin is carried by quarks. This has triggered a great deal of interest in the subject. The other polarized structure function $g_2$ has no interpretation in purely partonic language. It involves a twist-three operator and thus offers the first direct measurement of higher twist operator matrix elements [2]. Experiments that measure $g_2$ are currently being performed at DESY and SLAC.

We have initiated a program to compute $F_1$, $F_2$, $g_1$ and $g_2$ on the lattice [3]. For an earlier attempt to compute the unpolarized structure functions see [4]. The theoretical basis for such a calculation is the operator product expansion (OPE), which relates the moments of the structure functions to forward nucleon matrix elements of certain local operators. Where a parton model interpretation exists, it can be mapped onto an OPE analysis. Our calculation will be in the quenched approximation, where internal quark loops are neglected. In this paper we shall also neglect gluonic operators, which contribute only to higher order in the coupling constant expansion. For the unpolarized structure functions we then have for the leading twist contribution

$$
\begin{aligned}
2 \int_0^1 dx\, x^{n-1} F_1(x, Q^2) &= \sum_{f=u,d} c_{1,n}^{(f)}(\mu^2/Q^2, g(\mu))\, v_n^{(f)}(\mu), \\
\int_0^1 dx\, x^{n-2} F_2(x, Q^2) &= \sum_{f=u,d} c_{2,n}^{(f)}(\mu^2/Q^2, g(\mu))\, v_n^{(f)}(\mu)
\end{aligned}
\quad (1)
$$

for $n \geq 2$ (generally even $n$), where

$$
\frac{1}{2} \sum_{\vec{s}} \langle \vec{p}, \vec{s} | \mathcal{O}_{\{\mu_1 \cdots \mu_n\}}^{(f)} | \vec{p}, \vec{s} \rangle = 2 v_n^{(f)} [p_{\mu_1} \cdots p_{\mu_n} - \text{traces}],
$$

$$
\mathcal{O}_{\mu_1 \cdots \mu_n}^{(f)} = \left(\frac{i}{2}\right)^{n-1} \bar{\psi} \gamma_{\mu_1} \overleftrightarrow{D}_{\mu_2} \cdots \overleftrightarrow{D}_{\mu_n} \psi - \text{traces}
\quad (2)
$$

with $\psi = u(d)$ for $f = u(d)$, and

$$
\begin{aligned}
c_{1,n}^{(f)}(\mu^2/Q^2, g(\mu)) &= Q^{(f)2}(1 + g(\mu)^2 \bar{c}_{1,n}(\mu^2/Q^2, g(\mu))), \\
c_{2,n}^{(f)}(\mu^2/Q^2, g(\mu)) &= Q^{(f)2}(1 + g(\mu)^2 \bar{c}_{2,n}(\mu^2/Q^2, g(\mu))).
\end{aligned}
\quad (3)
$$

Here $\mu$ denotes the subtraction point, and $\{\cdots\}$ indicates symmetrization. We have chosen the normalization $\langle \vec{p}, \vec{s} | \vec{p}', \vec{s}' \rangle = (2\pi)^3 2 E_{\vec{p}}\, \delta(\vec{p} - \vec{p}')\, \delta_{\vec{s}, \vec{s}'}$, $s^2 = -m_N^2$. The moments of $F_1$, $F_2$ have the parton model interpretation

$$
v_n^{(f)} = \langle x^{n-1} \rangle^{(f)}, \quad (4)
$$

where $x$ is the fraction of the nucleon momentum carried by the quarks. In the quenched approximation the above-mentioned equations hold for odd $n$ as well.

For the polarized structure functions we have, again for the leading twist contribution,

$$2 \int_0^1 dx\, x^n g_1(x, Q^2) = \frac{1}{2} \sum_{f=u,d} e_{1,n}^{(f)}(\mu^2/Q^2, g(\mu))\, a_n^{(f)}(\mu), \tag{5}$$

$$2 \int_0^1 dx\, x^n g_2(x, Q^2) = \frac{1}{2} \frac{n}{n+1} \sum_{f=u,d} [e_{2,n}^{(f)}(\mu^2/Q^2, g(\mu))\, d_n^{(f)}(\mu) - e_{1,n}^{(f)}(\mu^2/Q^2, g(\mu))\, a_n^{(f)}(\mu)] \tag{6}$$

for even $n$ and $n \geq 0$ ($n \geq 2$) for $g_1$ ($g_2$), where [2]

$$\begin{aligned}
\langle \vec{p}, \vec{s} | \mathcal{O}^{5(f)}_{\{\sigma\mu_1\cdots\mu_n\}} | \vec{p}, \vec{s} \rangle &= \frac{1}{n+1} a_n^{(f)} [s_\sigma p_{\mu_1} \cdots p_{\mu_n} + \cdots - \text{traces}], \\
\langle \vec{p}, \vec{s} | \mathcal{O}^{5(f)}_{[\sigma\{\mu_1]\cdots\mu_n\}} | \vec{p}, \vec{s} \rangle &= \frac{1}{n+1} d_n^{(f)} [(s_\sigma p_{\mu_1} - s_{\mu_1} p_\sigma) p_{\mu_2} \cdots p_{\mu_n} + \cdots - \text{traces}], \\
\mathcal{O}^{5(f)}_{\sigma\mu_1\cdots\mu_n} &= \left(\frac{i}{2}\right)^n \bar{\psi} \gamma_\sigma \gamma_5 \overleftrightarrow{D}_{\mu_1} \cdots \overleftrightarrow{D}_{\mu_n} \psi - \text{traces}
\end{aligned} \tag{7}$$

and

$$\begin{aligned}
e_{1,n}^{(f)}(\mu^2/Q^2, g(\mu)) &= Q^{(f)2}(1 + g(\mu)^2 \bar{e}_{1,n}(\mu^2/Q^2, g(\mu))), \\
e_{2,n}^{(f)}(\mu^2/Q^2, g(\mu)) &= Q^{(f)2}(1 + g(\mu)^2 \bar{e}_{2,n}(\mu^2/Q^2, g(\mu))).
\end{aligned} \tag{8}$$

In eq. (7) $[\cdots]$ indicates antisymmetrization. In parton model language

$$a_0^{(u)} = 2\Delta u, \; a_0^{(d)} = 2\Delta d, \tag{9}$$

where $\Delta u, \Delta d$ determine the fraction of the nucleon spin that is carried by the quarks. A similar interpretation holds for the higher spin operators. The structure function $g_2$ consists of two contributions: $a_n^{(f)}$ is the so-called Wandzura-Wilczek contribution [5] which corresponds to twist two, whereas $d_n^{(f)}$ corresponds to twist three. In addition to $d_n^{(f)}$ there is a contribution to $g_2$ from a twist-three operator which is proportional to the quark mass [6]. Since we are mainly interested in the chiral limit we have neglected that.

## 2 Lattice Calculation

We perform our quenched QCD calculations for Wilson fermions with $r = 1$ on a $16^3 \times 32$ lattice. For the gauge coupling we take $\beta \equiv 6/g^2 = 6.0$. To be able to extrapolate our results to the chiral limit, we run at three different hopping parameters, $\kappa = 0.155, 0.153$ and $0.1515$. This corresponds to physical quark masses $m_q$ of roughly 70, 130 and 190 MeV, respectively, using the perturbative relation

$$m_q a = 0.56 \left(\frac{1}{\kappa} - \frac{1}{\kappa_c}\right), \tag{10}$$

where $a$ is the lattice spacing.

For the gauge update we use a cycle consisting of a single three-hit Metropolis sweep followed by 16 overrelaxation sweeps [7]. We repeat this cycle 50 times in order to generate a new configuration. The resulting configurations seem to be independent. We see no correlations between hadronic quantities calculated on different gauge field configurations. The calculations are carried out on Quadrics Q16 and QH2 parallel computers. For details of the implementation of our code on these machines see [3]. So far we have collected of the order 1000, 600 and 400 independent configurations at the three $\kappa$ values.

To calculate the nucleon matrix elements we first compute two- and three-point correlation functions defined by

$$\begin{aligned} C_\Gamma(t,\vec{p}) &= \sum_{\alpha,\beta} \Gamma_{\beta,\alpha} \langle B_\alpha(t,\vec{p}) \bar{B}_\beta(0,\vec{p}) \rangle, \\ C_\Gamma(t,\tau,\vec{p},\mathcal{O}) &= \sum_{\alpha,\beta} \Gamma_{\beta,\alpha} \langle B_\alpha(t,\vec{p}) \mathcal{O}(\tau) \bar{B}_\beta(0,\vec{p}) \rangle. \end{aligned} \quad (11)$$

The lattice operators $\mathcal{O}$ used here are obtained from the operators in the euclidean continuum, which look exactly like the operators in eqs. (2), (7) up to factors of i, by replacing the covariant derivative by the lattice covariant derivative so that $\vec{D}_\mu$ becomes

$$\vec{D}_\mu(x,y) = \frac{1}{2}[U_\mu(x)\delta_{y,x+\hat{\mu}} - U_\mu^\dagger(x-\hat{\mu})\delta_{y,x-\hat{\mu}}]. \quad (12)$$

We write the ratio of three- to two-point correlation functions as

$$\begin{aligned} R(t,\tau,\vec{p},\Gamma,\mathcal{O}) &= C_\Gamma(t,\tau,\vec{p},\mathcal{O})/C_{\frac{1}{2}(1+\gamma_4)}(t,\vec{p}) \\ &= \frac{1}{2\kappa}\frac{E_{\vec{p}}}{E_{\vec{p}}+m_N} F(\Gamma,\mathcal{J}) \end{aligned} \quad (13)$$

(for $0 \ll \tau \ll t$) with

$$\begin{aligned} F(\Gamma,\mathcal{J}) &= \frac{1}{4}\text{Tr}\left[\Gamma N \mathcal{J} N\right], \\ N &= \gamma_4 - i\vec{p}\vec{\gamma}/E_{\vec{p}} + m_N/E_{\vec{p}} \end{aligned} \quad (14)$$

and $\mathcal{J}$ defined by

$$\langle \vec{p},\vec{s}|\mathcal{O}|\vec{p},\vec{s}\rangle = \bar{u}(\vec{p},\vec{s})\mathcal{J} u(\vec{p},\vec{s}). \quad (15)$$

When calculating three-point functions it is particularly important that the baryon operator $B$ has only little overlap with excited baryon states, in order to make the plateau region in $\tau$ as broad as possible. As our basic proton operator we use (with $C = \gamma_4\gamma_2$ in our representation)

$$B_\alpha(t,\vec{p}) = \sum_{\vec{x},a,b,c} e^{-i\vec{p}\vec{x}} \epsilon_{abc} u_\alpha^a(x)(u^b(x) C \gamma_5 d^c(x)) \quad (16)$$

with two important improvements. First we use 'Jacobi smearing' [8] (a version of [9]) in order to have an extended proton operator. Thus each quark operator in eq. (11) is replaced by

$$\psi \to \psi^S = \sum_{n=0}^{N_s} (\kappa_s \vec{D})^n \psi, \quad (17)$$

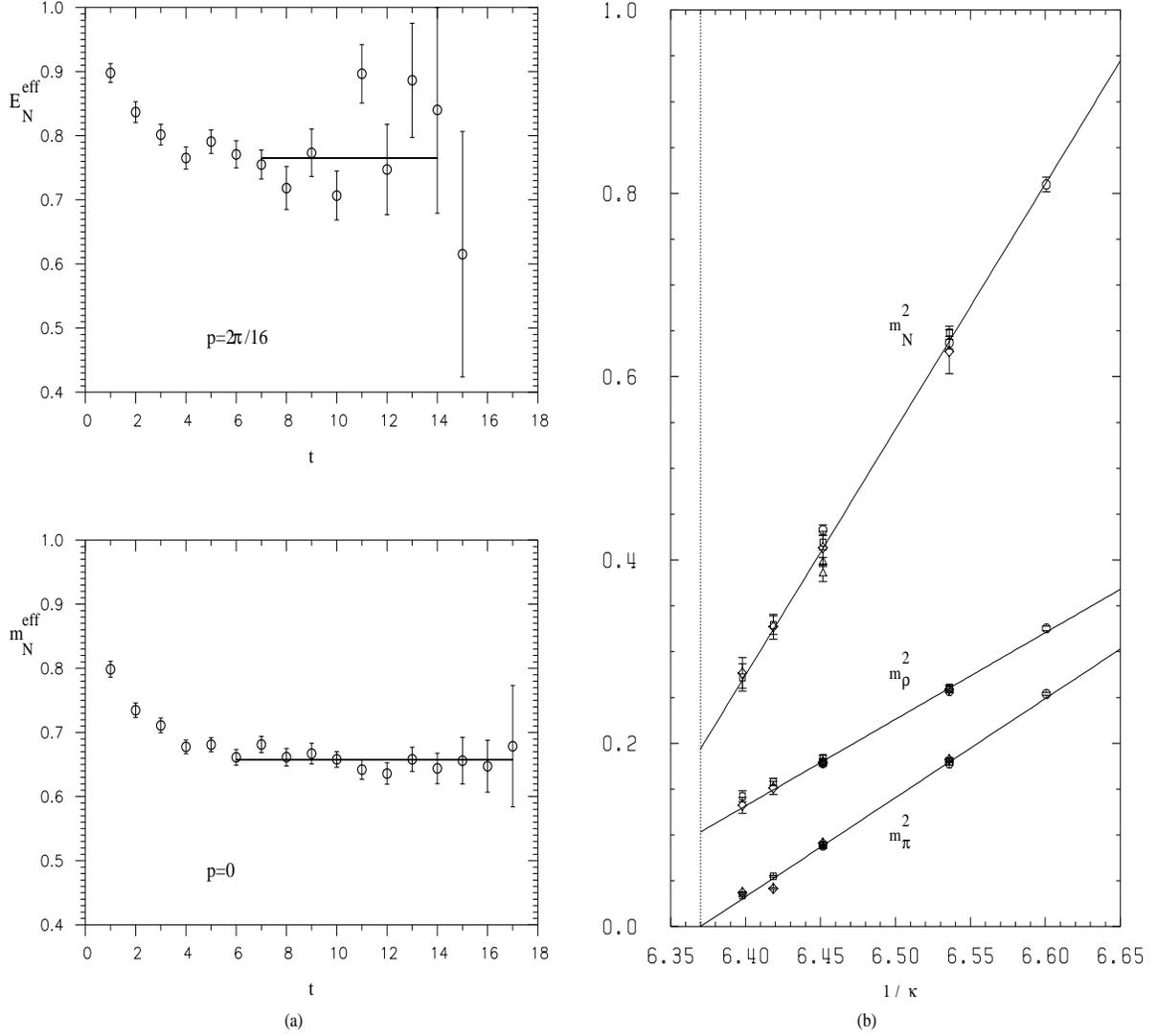

Figure 1: (a) Effective nucleon mass plot for $\vec{p} = 0$ (bottom), and effective nucleon energy plot for $|\vec{p}| = 2\pi/16$ (top) at $\kappa = 0.155$. Both source and sink are smeared. The horizontal lines indicate the result of the fit as well as the fit interval. (b) $m_\pi^2$, $m_\rho^2$ and $m_N^2$ as a function of $1/\kappa$ together with other recent results of the literature: ◯ this work, □ Ref. [10], ◇ Ref. [11], △ Ref. [12]. The fit to all data points combined gives $\kappa_c = 0.15699(5)$.

|       | $\kappa$ |       |       |
|-------|----------|-------|-------|
|       | 0.1515   | 0.153 | 0.155 |
| $m_\pi$ | 0.504(2) | 0.422(2) | 0.297(2) |
| $m_\rho$ | 0.570(2) | 0.507(2) | 0.422(2) |
| $m_N$ | 0.900(5) | 0.798(5) | 0.658(5) |

Table 1: The hadron masses in lattice units at $\beta = 6.0$.

and similarly for $\bar\psi$ as we smear both source and sink. We found suitable values of the parameters to be $N_s = 50, \kappa_s = 0.21$, which for our largest $\kappa$ value gave a *rms* radius of about 4, corresponding roughly to 0.5 fm, i.e. half the nucleon radius. Secondly we replace each spinor by

$$\psi \to \psi^{NR} = \tfrac{1}{2}(1 + \gamma_4)\,\psi, \quad \bar\psi \to \bar\psi^{NR} = \bar\psi\,\tfrac{1}{2}(1 + \gamma_4). \tag{18}$$

This replacement leaves quantum numbers unchanged, but we expect it to improve overlap with those baryons which have slow-moving valence quarks. Practically this means that for each baryon propagator we invert on a smeared local source and consider only the first two Dirac components. So we only have $2 \times 3$ inversions to perform rather than the usual $4 \times 3$ inversions, which saves a factor two in computer time in the inversion.

In Ref. [3] we have seen that the projection (18) is particularly effective at reducing unwanted backward propagating states, which extends the window that one can practically use for matrix element calculations to well above half the temporal extent of the lattice. In Fig. 1a we show a plot of the resulting effective nucleon energy, as given by $\ln(C(t)/C(t+1))$, for $\vec{p} = 0$ and $|\vec{p}| = 2\pi/16$ at our smallest quark mass. For zero nucleon momentum we find a good plateau with a proton mass of 0.658(5). For the lowest non-zero momentum we find an energy of 0.765(11), which is in good agreement with the continuum dispersion relation. In both cases we see that after a distance of about four time units there is very little trace of an excited state. In Table 1 we give the mass values of the nucleon together with those of the $\pi$ and $\rho$ for our three values of $\kappa$. The chiral limit is obtained by extrapolating in $1/\kappa$ to zero $\pi$ mass. Assuming, as usual, that $m_\pi^2$ depends linearly on $1/\kappa$, we obtain from our data the critical value $\kappa_c = 0.15693(4)$. In Fig. 1b we plot $m_\pi^2$, $m_\rho^2$ and $m_N^2$ as a function of $1/\kappa$. In this plot we have also included other recent results at $\beta = 6.0$. A combined fit gives $m_N/m_\rho = 1.37$ in the chiral limit.

To calculate three-point functions we require additional propagators, one for each chosen $t$, $\vec{p}$ and $\Gamma$. We have fixed $t$ at 13 and have chosen $\Gamma = \tfrac{1}{2}(1 + \gamma_4)$, corresponding to the unpolarized case, and $\Gamma = \tfrac{1}{2}(1 + \gamma_4)i\gamma_5\gamma_2$, corresponding to polarization $(+\ -\ -)$ in the 2-direction. For the momentum we have taken $\vec{p} = 0$ and $\vec{p} = (2\pi/16, 0, 0) \equiv (p_1, 0, 0)$. We have also considered $\psi = u, d$ separately. This means that we must find $2 \times 2 \times 2 = 8$ (half) quark propagators. The choice $t = 13$ is sufficient. Larger values of $t$ lead to unacceptably large errors in the signal for $R$. Test runs for $t = 17$ turned out to have $O(2)$ larger errors, which roughly corresponds to the increase in the noise in the baryon correlation function from $t = 13$ to $t = 17$.

| $\langle \mathcal{O} \rangle$ | Components | Representation | | |
|---|---|---|---|---|
| | | Ref. [13] | Ref. [14] | $C$ |
| $v_{2,a}$ | $\mathcal{O}_{\{14\}}$ | $\tau_3^{(6)}$ | $6^{(+)}$ | $+$ |
| $v_{2,b}$ | $\mathcal{O}_{\{44\}} - \frac{1}{3}(\mathcal{O}_{\{11\}} + \mathcal{O}_{\{22\}} + \mathcal{O}_{\{33\}})$ | $\tau_1^{(3)}$ | 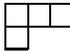 $^{(+)}$ | $+$ |
| $v_3$ | $\mathcal{O}_{\{114\}} - \frac{1}{2}(\mathcal{O}_{\{224\}} + \mathcal{O}_{\{334\}})$ | $\tau_1^{(8)}$ | $8^{(+)}$ | $-$ |
| $v_4$ | $\mathcal{O}_{\{1144\}} + \mathcal{O}_{\{2233\}} - \mathcal{O}_{\{1133\}} - \mathcal{O}_{\{2244\}}$ | $\tau_1^{(2)}$ | 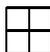 $^{(+)}$ | $+$ |
| $a_0$ | $\mathcal{O}_2^5$ | $\tau_4^{(4)}$ | $(\frac{1}{2},\frac{1}{2})^{(-)}$ | $+$ |
| $a_2$ | $\mathcal{O}_{\{214\}}^5$ | $\tau_3^{(4)}$ | $(\overline{\frac{1}{2},\frac{1}{2}})^{(+)}$ | $+$ |
| $d_2$ | $\mathcal{O}_{[2\{1]4\}}^5$ | $\tau_1^{(8)}$ | $8^{(+)}$ | $+$ |

Table 2: The lattice operators and their representation. The momentum is taken to be $\vec{p} = (2\pi/16, 0, 0) \equiv (p_1, 0, 0)$ in the case of $v_{2,a}, v_3, v_4, a_2, d_2$ and $\vec{p} = 0$ elsewhere. $C$ denotes charge conjugation.

## 3 Lattice Operators and their Renormalization

The bare lattice operators, $\mathcal{O}(a)$, are in general divergent. We define finite operators $\mathcal{O}(\mu)$, renormalized at the scale $\mu$, by

$$\mathcal{O}(\mu) = Z_{\mathcal{O}}((a\mu)^2, g(a))\mathcal{O}(a), \qquad (19)$$

where

$$\langle q(p)|\mathcal{O}(\mu)|q(p)\rangle = \langle q(p)|\mathcal{O}(a)|q(p)\rangle \mid_{p^2=\mu^2}^{tree} \qquad (20)$$

with $|q(p)\rangle$ being a quark state of momentum $p$. In the limit $a \to 0$ this definition amounts to the continuum, momentum subtraction renormalization scheme.

The lattice operators transform under the discrete hypercubic group $H(4)$ [13, 14]. They must be constructed such that they belong to a definite irreducible representation of the latter. In particular they must not mix with lower-dimensional operators. This is prerequisite to the operators being multiplicatively renormalizable. Furthermore, from the more practical point of view, the operators should only require a non-zero spatial momentum in at most one direction. We have considered the operators listed in Table 2. For the group theoretical classification of the lattice operators see Ref. [15]. The calculation of $v_{2,a}$, $v_3$, $v_4$, $a_2$ and $d_2$ requires non-vanishing nucleon momenta. Note that for the quenched theory there is no mixing with gluon operators. In the continuum limit the matrix elements $v_{2,a}$ and $v_{2,b}$ should be equal. At finite lattice spacing this provides us with a consistency check and gives information about possible lattice artifacts.

We have computed the renormalization constants for our operators in the quenched approximation for Wilson fermions in perturbation theory to one loop order. For this task we have developed packages of computer algebraic programs using $Mathematica$ and $Maple$ to such a level that all what is needed as input is to state the Feynman rules in symbolic form, both for the continuum and the lattice part of the calculation. We will summarize our results here. A detailed account of our calculation will be given elsewhere [16].

| $\langle\mathcal{O}\rangle$ | | $\gamma_\mathcal{O}$ | $B_\mathcal{O}$ | $Z_\mathcal{O}(1, g = 1.0)$ | $B_\mathcal{O}^{\overline{MS}}$ |
|---|---|---|---|---|---|
| $v_{2,a}$ | | $\frac{16}{3}$ | -3.165(6) | 1.0267(1) | $-\frac{40}{9}$ |
| $v_{2,b}$ | | $\frac{16}{3}$ | -1.892(6) | 1.0160(1) | $-\frac{40}{9}$ |
| $v_3$ | {}{}: | $\frac{25}{3}$ | -19.572(10) | 1.1653(1) | $-\frac{67}{9}$ |
| | {}(): | 0 | 0.370(10) | -0.0031(1) | |
| $v_4$ | | $\frac{157}{15}$ | -37.16(30) | 1.314(3) | $-\frac{2216}{225}$ |
| $a_0$ | | 0 | 15.795(3) | 0.8666(0) | 0 |
| $a_2$ | | $\frac{25}{3}$ | -19.560(10) | 1.1652(1) | $-\frac{67}{9}$ |
| $d_2$ | | $\frac{7}{3}$ | -15.680(10) | 1.1324(1) | $-\frac{13}{12}$ |

Table 3: The renormalization constants in the quenched approximation. The errors quoted are a conservative estimate of the uncertainties in the numerical evaluation of the integrals involved. The numbers in the rightmost column represent the contribution of the continuum operators computed in the $\overline{MS}$ scheme.

In the case of $v_3$ it turns out that the operator $\mathcal{O}_{\{114\}} - \frac{1}{2}(\mathcal{O}_{\{224\}} + \mathcal{O}_{\{334\}})$ mixes with the operator [15]

$$\mathcal{O}_{(114)} - \frac{1}{2}(\mathcal{O}_{(224)} + \mathcal{O}_{(334)}), \quad \mathcal{O}_{(\mu\mu\nu)} = \mathcal{O}_{\mu\mu\nu} + \mathcal{O}_{\mu\nu\mu} - 2\mathcal{O}_{\nu\mu\mu} \qquad (21)$$

under renormalization. This operator is of mixed symmetry, is traceless and corresponds to the representation $8^{(+)}, C = -$ as well. Thus we have

$$\mathcal{O}_{\{\}}(\mu) = Z_{\{\}\{\}}\mathcal{O}_{\{\}}(a) + Z_{\{\}()}\mathcal{O}_{()}(a), \qquad (22)$$

where we have used a short-hand notation for the operators in Table 2 and eq. (21). We write $(C_F = 4/3)$

$$Z_\mathcal{O}((a\mu)^2, g) = \mathbb{1} - \frac{g^2}{16\pi^2}C_F[\gamma_\mathcal{O}\ln(a\mu) + B_\mathcal{O}]. \qquad (23)$$

This is to be interpreted as a matrix equation in the case of $v_3$. Our results for the anomalous dimensions $\gamma_\mathcal{O}$ and the $B_\mathcal{O}$'s are given in Table 3 for $r = 1$. The renormalization constants $Z_{v_{2,b}}$ and $Z_{a_0}$ have been given before in the literature [17, 18, 19]. We agree with the results of these authors. In the case of $v_3$ the off-diagonal component of $Z$ is negligibly small.

The structure functions do not depend on $\mu$, but $\langle x^{n-1}\rangle$ and $\Delta u$, $\Delta d$ do. In the following we shall quote our results for

$$\mu^2 = Q^2 = a^{-2} \approx 2\text{GeV}^2, \qquad (24)$$

which eliminates the logarithms in the Wilson coefficients and renormalization constants, and we will denote $Z_{\mathcal{O}}(1, g = 1.0)$ by $Z_{\mathcal{O}}$. The corresponding numerical values are also listed in Table 3. As the Wilson coefficients are generally computed in the $\overline{MS}$ regularization scheme, one needs to know the renormalization constants in this scheme too. In Table 3 we state the contribution of the continuum operators computed in the $\overline{MS}$ scheme. The difference of the $B_{\mathcal{O}}$'s then gives the result in the $\overline{MS}$ scheme.

The renormalization constants receive contributions from five different types of diagrams: the vertex, the leg self-energy, the leg tadpole, the operator tadpole and the operator comb diagrams. The tadpole diagrams give by far the largest contribution to the renormalization constants. The leg tadpole contribution is the same for all operators. The operator tadpole contribution is proportional to the number of covariant derivatives and has opposite sign to the leg tadpole. Leg and operator tadpole diagrams cancel each other in $v_2$. This accounts for the small values of $B_{v_{2,a}}, B_{v_{2,b}}$. In $a_0$ only the leg tadpole contributes. In all other cases it is the operator tadpole diagram which dominates.

## 4   Structure Function Results

The next step is to calculate the ratio of three- to two-point correlation functions $R$, as given in eq. (13), for the operators listed in Table 2. To make sure that we are computing the matrix elements of the lowest-lying state, i.e. the nucleon, we must look for plateaus in $\tau$, the time distance of the operator from the source, for $0 \ll \tau \ll t = 13$. In Fig. 2 we show $R$ as a function of $\tau$ for six of our operators at $\kappa = 0.153$. The ratio $R_{v_{2,b}}$ not shown here is of the same quality as $R_{v_{2,a}}$. We find in all cases that the signal is practically constant for time distances larger than two lattice spacings from the source and from the sink. For $13 \leq \tau$ the signal is practically zero as one would expect. The fit interval is taken to be $4 \leq \tau \leq 9$. The result of the fit is shown by the horizontal lines, and the errors are indicated by the dotted lines.

The renormalized operator matrix elements are obtained from the ratio $R$ by

$$\begin{aligned}
R_{v_{2,a}} &= \frac{i}{Z_{v_{2,a}}} \frac{1}{2\kappa} p_1 v_{2,a}, & R_{v_{2,b}} &= -\frac{1}{Z_{v_{2,b}}} \frac{1}{2\kappa} m_N v_{2,b}, \\
R_{v_3} &= -\frac{1}{Z_{v_3}} \frac{1}{2\kappa} p_1^2 v_3, & R_{v_4} &= \frac{1}{Z_{v_4}} \frac{1}{2\kappa} E_{p_1} p_1^2 v_4, \\
R_{a_0} &= \frac{i}{Z_{a_0}} \frac{1}{2\kappa} \frac{m_N}{2E_{p_1}} a_0, & R_{a_2} &= \frac{1}{Z_{a_2}} \frac{1}{2\kappa} \frac{1}{6} m_N p_1 a_2, \\
R_{d_2} &= \frac{1}{Z_{d_2}} \frac{1}{2\kappa} \frac{1}{3} m_N p_1 d_2.
\end{aligned} \qquad (25)$$

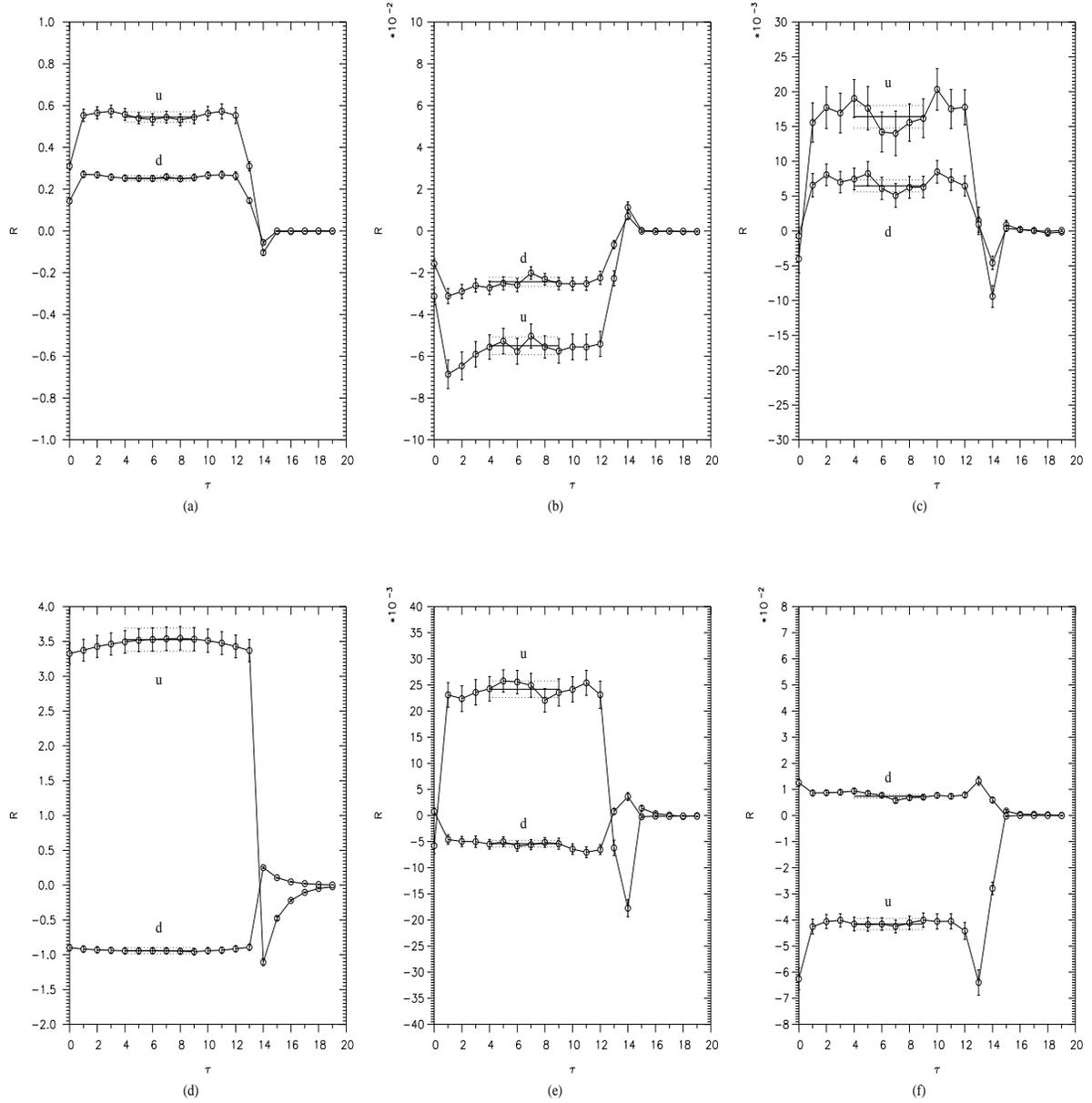

Figure 2: The ratio $R$ for $u$ and $d$ quark insertions for $\kappa = 0.153$. (a) $-iR_{v_{2,a}}$, (b) $R_{v_3}$, (c) $R_{v_4}$, (d) $-iR_{a_0}$, (e) $R_{a_2}$ and (f) $R_{d_2}$. Both source and sink are smeared. The source is at $t = 0$, the sink at $t = 13$.

We have defined the continuum quark fields by $\sqrt{2\kappa}$ times the lattice quark fields. For the renormalization constants we take the perturbative values given in Table 3 and $Z_{v_4} = 1$. In the case of $v_3$ we have also computed the nucleon matrix element of the operator in eq. (21). It turned out to be noisy and consistent with zero within an error of roughly $1/5$ the magnitude of the leading symmetric contribution. Given the small off-diagonal component of the renormalization constant, we may thus safely neglect the effect of mixing. Tadpole resummation [20] would leave $Z_{v_{2,a}}$, $Z_{v_{2,b}}$ unchanged, while it would change the other renormalization constants by a few percent. The exact amount depends on how it is implemented, and there is considerable freedom to do so. (Better is to compute the renormalization constants non-perturbatively [21] what we are doing now [22].) The results are plotted in Fig. 3, and the numerical values are listed in Table 4. All our results are given for the proton. The distribution functions of the neutron are obtained by interchanging $u$ and $d$.

We shall now discuss our results in detail. The first important observation to make is that the values of $\langle x \rangle_a$ and $\langle x \rangle_b$, which are obtained from different representations of the hypercubic group $H(4)$ (cf. Table 2), are consistent with each other, within the error bars. This indicates that lattice artifacts are presumably small. A second observation is that all matrix elements show roughly a linear behavior in $1/\kappa$, i.e. in the quark mass (cf. eq. (10)). The lines shown in Fig. 3 are linear fits to the data. The result of the extrapolation is indicated by the solid circles and the solid box, and the numerical values of the fit are given in the last column of Table 4.

Let us concentrate on the moments of the unpolarized structure functions first. We see that the lowest moment ($n = 2$) is practically independent of the quark mass. For growing $n$ the moments show a stronger and stronger increase with the quark mass. For the distribution function itself this means that at small $x$ quark mass effects are negligible, while at intermediate and large $x$ its shape depends strongly on the magnitude of the quark mass. In the limit of large quark masses the higher moments approach the predictions of the non-relativistic quark model. In particular we find $\langle x^{n-1} \rangle^{(u)} \approx 2 \langle x^{n-1} \rangle^{(d)}$ for all $n$. In the chiral limit the picture changes completely. Whereas at small $x$ the ratio of $u$ to $d$ distribution is roughly two, the ratio increases rapidly for larger values of $x$.

We may compare our results in the chiral limit with the phenomenological valence quark distribution functions. In Fig. 3a-c we show the results of a recent such fit [23]. For $n = 2$ the lattice values are significantly larger than the phenomenological values, for $n = 3$ they are about equal within the statistical errors, and for $n = 4$ the lattice values are significantly smaller than the phenomenological values. This holds for both, $u$ and $d$ quark distributions. Thus, our calculation predicts a valence quark distribution function that is more singular at small $x$ than the phenomenological one, i.e. has a somewhat larger Regge intercept. At the moment we have no explanation for this discrepancy. According to our calculation 64% of the nucleon's momentum is carried by the (valence) quarks.

We shall now turn to the discussion of the polarized structure functions. Let us first focus on $\Delta u$ and $\Delta d$, which in the quenched approximation determine the fraction of the proton spin that is carried by the valence quarks. Sea quark effects may be neglected for

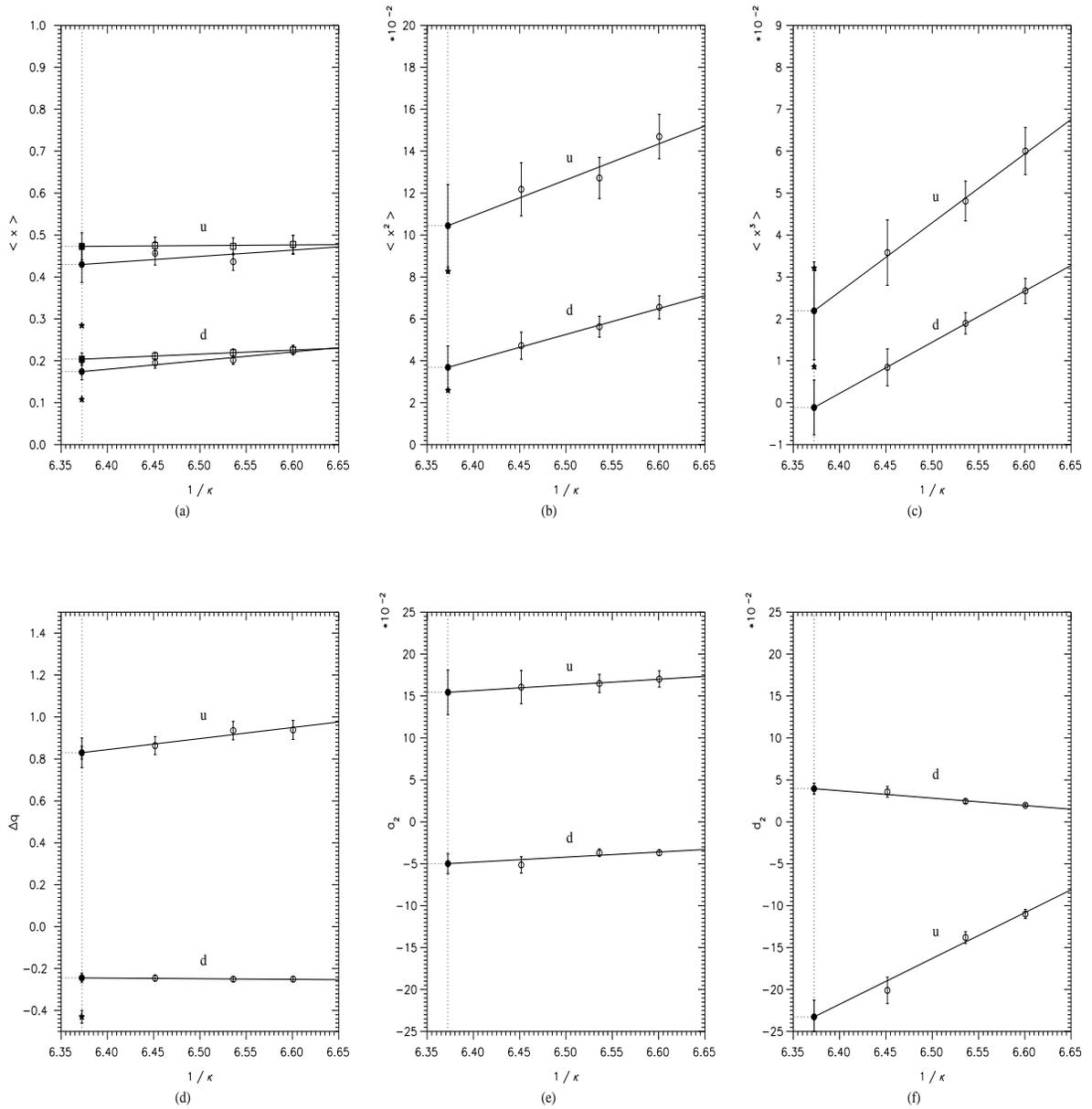

Figure 3: The moments of the proton structure functions as a function of $1/\kappa$, together with a linear fit to the data. The solid symbols indicate the extrapolation to the chiral limit. In (a) circles refer to $\langle x \rangle_a$, boxes to $\langle x \rangle_b$. In (a-c) we compare our results with the phenomenological valence quark distribution of Ref. [23] (fit $D_-$). The phenomenological moments are marked by asterisks. In (d) we compare our numbers with the phenomenological values of Ref. [24], which are marked by asterisks as well. (The value for $\Delta u$ is hidden behind the lattice number.)

| Observable | $\kappa$ | | | |
|---|---|---|---|---|
| | 0.1515 | 0.153 | 0.155 | $\kappa_c = 0.1569$ |
| $\langle x \rangle_a^{(u)}$ | 0.477(23) | 0.436(20) | 0.457(28) | 0.430(43) |
| $\langle x \rangle_b^{(u)}$ | 0.478(22) | 0.473(20) | 0.475(20) | 0.473(32) |
| $\langle x \rangle_{av.}^{(u)}$ | 0.478(16) | 0.455(14) | 0.466(16) | 0.452(26) |
| $\langle x \rangle_a^{(d)}$ | 0.226(11) | 0.201(10) | 0.195(12) | 0.174(20) |
| $\langle x \rangle_b^{(d)}$ | 0.226(10) | 0.219(09) | 0.211(09) | 0.204(15) |
| $\langle x \rangle_{av.}^{(d)}$ | 0.226(08) | 0.210(07) | 0.203(07) | 0.189(12) |
| $\langle x^2 \rangle^{(u)}$ | 0.147(11) | 0.127(10) | 0.122(13) | 0.104(20) |
| $\langle x^2 \rangle^{(d)}$ | 0.066(05) | 0.056(05) | 0.047(06) | 0.037(10) |
| $\langle x^3 \rangle^{(u)}$ | 0.060(05) | 0.049(05) | 0.035(08) | 0.022(11) |
| $\langle x^3 \rangle^{(d)}$ | 0.026(03) | 0.018(03) | 0.008(04) | -0.001(07) |
| $\Delta u$ | 0.938(45) | 0.935(44) | 0.863(43) | 0.830(70) |
| $\Delta d$ | -0.250(12) | -0.250(12) | -0.246(14) | -0.244(22) |
| $a_2^{(u)}$ | 0.170(10) | 0.165(11) | 0.161(20) | 0.154(27) |
| $a_2^{(d)}$ | -0.037(03) | -0.037(04) | -0.051(10) | -0.050(12) |
| $d_2^{(u)}$ | -0.110(05) | -0.138(07) | -0.201(16) | -0.233(20) |
| $d_2^{(d)}$ | 0.020(02) | 0.024(02) | 0.036(06) | 0.040(07) |

Table 4: Structure function results for the proton. All numbers refer to the momentum subtraction scheme.

heavy quarks, and they drop out in the difference $\Delta u - \Delta d$. In the chiral limit we obtain

$$\Delta u - \Delta d \equiv g_A = 1.07(9). \tag{26}$$

This is to be compared with the experimental value of the axial vector coupling constant $g_A = 1.26$. In Fig. 3d we compare $\Delta u$ and $\Delta d$ individually with a recent phenomenological fit of the polarized structure function data [24], which naturally includes sea quark effects. If we add the sea quark contribution to our results – a recent lattice calculation [25] finds $\Delta \bar{u} = \Delta \bar{d} = -0.14(5)$, $\Delta \bar{s} = -0.13(4)$ using perturbative renormalization factors – we would favor a somewhat smaller value for $\Delta u$ than the fitted value. For the total quark spin contribution to the nucleon spin we would furthermore obtain $\Delta \Sigma = 0.18(8)$, in agreement with the result of a full QCD calculation [26], i.e. including dynamical quarks. For heavy quark masses we find $\Delta u \approx 1$ and $\Delta d \approx -1/4$, in good agreement with the three-quark model [27].

By comparing the moments $a_0 = 2\Delta q$ and $a_2$ with those of the unpolarized structure functions we find that in the chiral limit $g_1$ is less singular than $F_1$ as $x$ goes to zero. This is also what one finds experimentally [28]. In the limit of large quark masses, on the other hand, it seems that $g_1$ is proportional to $F_1$.

If we combine our results with the perturbatively known [29] Wilson coefficients we can compute the moments of $g_1$. In the chiral limit we obtain for the lowest moment

$$\int_0^1 dx\, g_1(x, Q^2) = \begin{cases} 0.166(16) & \text{proton}, \\ -0.008(09) & \text{neutron}. \end{cases} \quad (27)$$

Remember that $Q^2 \approx 2\,\text{GeV}^2$. For the difference of proton and neutron structure functions we find

$$\int_0^1 dx\, (g_1^p(x, Q^2) - g_1^n(x, Q^2)) = 0.174(15). \quad (28)$$

This is the quantity which should be compared with experiment, because here the sea quark contribution drops out. Our result is in good agreement with the phenomenological analysis [24, 28]. In the higher moments of $g_1$ sea quark effects should not play any role anymore either. In the chiral limit we obtain

$$\int_0^1 dx\, x^2 g_1(x, Q^2) = \begin{cases} 0.0150(32) & \text{proton}, \\ -0.0012(20) & \text{neutron}. \end{cases} \quad (29)$$

Here we have converted the renormalization constants to the $\overline{MS}$ scheme, because the Wilson coefficients were computed in this scheme too. This result is consistent with experiment [30].

Let us finally discuss the structure function $g_2$. From Fig. 3f we read off that the twist-three contribution $d_2$ is strongly mass dependent. While $d_2$ approaches zero in the heavy quark limit, for both $u$ and $d$ quark insertion, it is of the same order of magnitude as its twist-two counterpart $a_2$ for small quark masses. In the chiral limit we obtain

$$\int_0^1 dx\, x^2 g_2(x, Q^2) = \begin{cases} -0.0161(16) - 0.0100(22) = -0.0261(38) & \text{proton}, \\ -0.0013(09) + 0.0009(13) = -0.0004(22) & \text{neutron}. \end{cases} \quad (30)$$

As before, Wilson coefficents [29] and renormalization constants are consistently computed in the $\overline{MS}$ scheme. In eq. (30) the first number comes from $d_2$, while the second number comes from $a_2$ (cf. eq. (6)). We see that the twist-three operator provides the dominant contribution. The Wandzura-Wilczek description of $g_2$ [5] is a valid approximation for large quark masses, but for light quark masses it is definitely not. Our results seem to be in disagreement with recent estimates based on sum rules [31], which suggest that for the proton $d_2$ is very small.

## 5 Conclusion

We have presented results of a calculation of the lower moments of the polarized and unpolarized deep-inelastic structure functions of the nucleon. The calculation has been performed in the quenched approximation, where sea quark effects are neglected, and it was done for three different quark masses. This allowed us to extrapolate our results to the chiral limit.

The valence quark distributions that we have obtained differ somewhat from the phenomenological ones [23]. One explanation could be that at smaller values of $Q^2$ higher twist contributions are non-negligible, which have not been included in the phenomenological analysis. We plan to investigate this possibility in the future. Our results for the polarized structure functions are consistent with experiment, as far as data exist. A surprise was that the twist-three operator contributed so much to $g_2$.

It was interesting to see how the results varied with the quark mass. At large quark masses our results agree largely with what one would expect on the basis of the quark model. For small quark masses there are, however, significant changes.

With the (raw) lattice data being relatively accurate now, the calculation of the renormalization constants has become a major issue. So far we have computed the renormalization constants in perturbation theory to one loop order. We hope to do better in the near future [22].

The renormalization constant for $v_3$ has independently been computed by the Rome group [32]. We have compared our results with theirs at intermediate stages of the calculation, and they agreed. These authors use a slightly different basis of operators from ours though.

# Acknowledgments

This work was supported in part by the Deutsche Forschungsgemeinschaft. The numerical calculations were performed on the Quadrics parallel computers at Bielefeld University and at DESY (Zeuthen). We wish to thank both institutions for their support and in particular the system managers M. Plagge and H. Simma for their help. We furthermore like to thank S. Capitani and G. Rossi for discussions on the problem of perturbative renormalization.